%%%%%%%%%%%%%%%%%%%%%%%%%%%%%%%%%%%%%%%%%%%%%%%%%%%%%
\magnification=\magstep1
\baselineskip=3ex
%\raggedbottom
\font\fivepoint=cmr5
\font\sixpoint=cmr6
\headline={\hfill{\fivepoint  EHLJY 27/Feb/98}}
\def\simt{\mathrel{\rlap{\hbox{$\sim$}}\raise.9ex\hbox{{\fivepoint
$\,$T}}}}
\def\sima{\mathrel{\rlap{\hbox{$\sim$}}\raise.95ex\hbox{{\fivepoint
$\,$A}}}}

\def\uprho{\raise1pt\hbox{$\rho$}}

\parindent =18pt
\parskip = 9pt

\input epsf
\input rotate

\voffset .5in

\centerline{\bf  A GUIDE TO ENTROPY AND }
\smallskip
\centerline{\bf THE SECOND LAW OF  THERMODYNAMICS }

%\bigskip \noindent 
%\halign{#\hfil\qquad\qquad\hfil&#\hfil\cr 
%Elliott H.
%Lieb\footnote{$^*$}{\sixpoint Work partially supported by U.S. National
%Science Foundation grant PHY95-13072A01.} & Jakob
%Yngvason\footnote{$^{**}$}{\sixpoint Work partially supported by the
%Adalsteinn Kristjansson Foundation, University of Iceland.}\cr {\it
%Departments of Mathematics and Physics} & {\it Institut f\"ur
%Theoretische Physik}\cr {\it Princeton University} & {\it Universit\"at
%Wien}\cr {\it Princeton, New Jersey  08544-0708} & {\it Boltzmanngasse
%5, A 1090 Vienna} \cr
%{\it U.S.A.}\footnote{}{\baselineskip=0.6\baselineskip\hskip
%-\parindent\sixpoint \copyright  1997 by the authors.  Reproduction of
%this article, by any means, is permitted for non-commercial
%purposes.\par} & {\it Austria} \cr } \bigskip

\bigskip
\centerline{Elliott H. Lieb\footnote{$^*$}{\sixpoint Work partially
supported by U.S. National Science Foundation grant PHY95-13072A01.} }
\centerline{\it Departments of  Mathematics and Physics, Princeton University}
\centerline{\it Jadwin Hall,  P.O. Box 708, Princeton, NJ  08544, USA}
\bigskip
\centerline{Jakob Yngvason
\footnote{$^{**}$}{\sixpoint Work partially
supported by the Adalsteinn Kristjansson Foundation,
University of Iceland.} }
\centerline{\it Institut f\"ur Theoretische Physik, Universit\"at Wien,}
\centerline{\it Boltzmanngasse 5, A 1090 Vienna, Austria}
\footnote{}{\baselineskip=0.6\baselineskip\hskip -\parindent\sixpoint
\copyright  1997 by the authors.
Reproduction of this article, by any means, is permitted for non-commercial
purposes.\par}

\bigskip

This article is intended for readers who, like us, were told that the
second law of thermodynamics is one of the major achievements of the
nineteenth century, that it is a logical, perfect and
unbreakable law --- but who were unsatisfied with the `derivations' of
the entropy principle as found in textbooks and in popular writings.  

A glance at the books will inform the reader that the law has `various
formulations' (which is a bit odd, as if to say the ten commandments
have various formulations) but they all lead to the existence of an
entropy function whose reason for existence is to tell us which
processes can occur and which cannot. We shall abuse language (or
reformulate it) by referring to the existence of entropy as {\it the}
second law. This, at least, is unambiguous.  The entropy we are talking
about is that defined by thermodynamics (and {\it not} some analytic
quantity, usually involving expressions such as $-p\ln p$, that appears
in information theory, probability theory and statistical mechanical
models).

There are three laws of thermodynamics (plus one more, due to Nernst,
which is mainly used in low temperature physics and is not immutable 
like the others). In
brief, these  are:

{\parindent=0pt
\hangindent=18pt \hangafter=0 {\it The Zeroth Law,} which expresses the
transitivity of equilibrium, and which is often said to imply the
existence of temperature as a parametrization of equilibrium states. We
use it below but formulate it without mentioning temperature. In fact,
temperature makes no appearance here until almost the very end.

\hangindent=18pt \hangafter=0 {\it The First Law,} which is conservation
of energy. It is a concept from mechanics  and provides the connection
between mechanics (and things like falling weights) and thermodynamics. 
We discuss this later on when we introduce simple   systems; the 
crucial usage of this law is that it allows   energy to be used
as one of the parameters describing the states of a simple system.

\hangindent=18pt \hangafter=0 {\it The Second Law.} Three popular
formulations of this law are:

\hangindent=25pt \hangafter=0 {\sl Clausius:\/} No process is possible,
the sole result of which is that heat is transferred from a body to a
hotter one.

\hangindent=25pt \hangafter=0 {\sl Kelvin (and Planck):\/} No process is
possible, the sole result of  which is that a body is cooled and work is
done.

\hangindent=25pt \hangafter=0 {\sl Carath\'eodory:\/} In any
neighborhood of any state there are states that cannot be reached from
it by an adiabatic process.

}

All three are supposed to lead to the entropy principle (defined
below).  These steps can be found in many books and will not be trodden
again here. Let us note in passing, however, that the first two use
concepts such as hot, cold, heat, cool that are intuitive but have to
be made precise before the statements are truly meaningful.  No one has
seen `heat', for example. The last (which uses the  term ``adiabatic
process", to be defined below) presupposes some kind of
parametrization of states by points in ${\bf R}^n$, and  the usual
derivation of entropy from it assumes some sort of differentiability;
such assumptions are beside the point as far as understanding the meaning of
entropy goes.

Why, one might ask, should a mathematician be interested in this matter,
which, historically, had something to do with  attempts to understand
and improve the  efficiency of steam engines? The answer, as we perceive
it, is that the law is really an interesting mathematical theorem about
orderings on sets, with profound physical implications. The axioms that
constitute this ordering are somewhat peculiar from the mathematical
point of view and might not arise in the ordinary  ruminations of
abstract thought. They are special, but important, and they are driven
by considerations about the world, which is what makes them so
interesting.  Maybe an ingenious reader will find an application of this
same logical structure to another field of science.

The basic input in our analysis is a certain kind of ordering on a set
and denoted by 
$$ \prec 
$$ 
(pronounced `precedes').  It is transitive and reflexive as in A1, A2
below, but $X\prec Y$ and $Y\prec X$ does not imply $X=Y$,  so it is a
`preorder'.  The big  question is whether $\prec$ can be encoded in an
ordinary, real-valued  function on the set, denoted by $S$, such that if
$X$ and $Y$ are related by $\prec$, then $S(X)\leq S(Y)$ if and only if
$X\prec Y$.  The function $S$ is also required to be additive and
extensive in a sense that will soon be made precise.

A helpful analogy is the question: When can a vector-field, $V(x)$, on
${\bf R}^3$ be encoded in an ordinary function, $f(x)$, whose gradient
is $V$? The well-known answer is that a necessary and sufficient
condition is that ${\rm curl}\, V=0$. Once $V$ is observed to
have this property one thing becomes evident and important: It is
necessary to measure the integral of $V$ only along some curves---not
all curves---in order to deduce the integral along {\it all} curves. The
encoding then has enormous predictive power about the nature of future
measurements of $V$.  In the same way, knowledge of the function
$S$ has enormous predictive power in the hands of chemists, engineers
and others concerned with the ways of the  physical  world.

Our concern will be the existence and properties of $S$, starting from
certain natural axioms about the relation $\prec$.  We present our
results without  proofs, but full details, and a discussion of related
previous work on the foundations of classical thermodynamics, are given
in [7]. The literature on this subject is extensive and it is not
possible to give even a brief account of it here, except for mentioning
that the previous work closest to ours is that of [6], and [2], (see
also [4], [5] and [9]). These other approaches are also based on an
investigation of the relation $\prec$, but the overlap with our work is
only partial.  In fact, a major part of our work is the derivation of a
certain property (the ``comparison hypothesis" below), which is taken as
an axiom in the other approaches.  It was a remarkable and largely
unsung  achievement of Giles [6] to realize the full power of this property.

{\parindent = 18pt
Let us begin the story with some basic concepts. 

\item{1.} {\it Thermodynamic System:} Physically, this consists of
certain specified amounts of certain kinds of matter, e.g., a gram of
hydrogen in a container with a piston, or a gram of hydrogen and a gram
of oxygen in two separate containers, or a gram of hydrogen and two
grams of hydrogen in separate containers.  The system can be in various
states which, physically, are {\it equilibrium states}.  The space of
states of the system is usually denoted by a symbol such as $\Gamma$
and states in $\Gamma$ by $X,Y,Z,$ etc.

 Physical motivation aside, a state-space, mathematically, is
just  a set --- to begin with; later on we will be interested in
embedding state-spaces in some convex subset of some $ {\bf R}^{n+1}$,
i.e., we will introduce coordinates. As we said earlier, however, the
entropy principle is quite independent of  coordinatization, 
Carath\'eodory's principle notwithstanding.

\item{2.} {\it Composition and scaling of states:}  The notion of
Cartesian product, $\Gamma_1 \times \Gamma_2$ 
%is self-evident, i.e., it
corresponds simply to the two (or more) systems being side by side on
the laboratory table; mathematically it is just another system (called a
{\it compound system}), and  we regard  the state space  $\Gamma_1
\times \Gamma_2$ as the same  as $\Gamma_2 \times \Gamma_1$. 
Points in $\Gamma_1 \times \Gamma_2$ are denoted by pairs $(X,Y)$, as
usual. The subsystems comprising a compound system are physically
independent systems, but they are allowed to interact with each other
for a period of time and thereby alter each other's state.

\item{} { The concept of  scaling is crucial. It is this concept that
makes our thermodynamics inappropriate for microscopic objects like
atoms or cosmic objects like stars.  For each state-space $\Gamma $ and
number $\lambda>0$ there is another state-space, denoted by
$\Gamma^{(\lambda)}$ with points denoted by $\lambda X$. This space is
called a {\it scaled copy} of $\Gamma$.  Of course we identify
$\Gamma^{(1)} =\Gamma$ and $1X=X$.  We also require
$(\Gamma^{(\lambda)})^{(\mu)} = \Gamma^{(\lambda\mu)}$ and $\mu(\lambda
X) = (\mu\lambda)X$.  The physical interpretation of
$\Gamma^{(\lambda)}$ when $\Gamma$ is the space of one gram of hydrogen,
is simply the state-space of $\lambda$ grams of hydrogen. The state
$\lambda X$ is the state  of  $\lambda $ grams of hydrogen with the 
same `intensive' properties as $X$, e.g., pressure, while `extensive'
properties like energy, volume etc.  are scaled by a factor $\lambda$,
(by definition).

 For any given $\Gamma$ we can form Cartesian product state  spaces
of the type $\Gamma^{(\lambda_1)} \times \Gamma^{(\lambda_2)} \times 
\cdots\times
\Gamma^{(\lambda_N)}$. These will be called {\it multiple scaled copies}
of $\Gamma$.

The notation $\Gamma^{(\lambda)}$ should be regarded as merely a
mnemonic at this point, but later on, with the embedding of $\Gamma$
into $ {\bf R}^{n+1}$, it will literally be $\lambda\Gamma= \{\lambda
X:X\in \Gamma\}$ in the usual sense.  

\item{3.} {\it Adiabatic accessibility:} Now we come to the ordering.
We say $X\prec Y$ (with $X$ and $Y$ {\it possibly 
in \underbar{different} state-spaces}) if  there  is an
{\it adiabatic process} that transforms $X$ into $Y$.

What does this  mean? Mathematically, we are just given a list of pairs 
$X \prec Y$. There is
nothing more to  be said, except that later on we will assume that this
list has certain properties that will lead to  interesting theorems
about this list, and will lead, in turn,  to the existence of an {\it
entropy function}, $S$ characterizing the list.

The physical interpretation is quite another matter. In text books a
process is usually called adiabatic if it takes place in `thermal
isolation`, which in turn means that 
`no heat is exchanged with the surroundings`.   
Such statements appear neither sufficiently general nor precise to us
and we prefer the following
version (which is in the spirit of Planck's formulation of the second
law [8]).
It has the great virtue (as discovered by Planck) that it avoids
having to distinguish between work and heat---or even having to define
the concept of heat.  We
emphasize, however, that  the theorems do not require 
agreement with our physical definition of adiabatic process; other
definitions are conceivably possible.

\item{} {
{\it A state $Y$ is adiabatically
accessible from a state $X$, in symbols $X\prec Y$, if it is possible to
change the state from $X$ to $Y$ by means of an interaction with some
device consisting  of some auxiliary system and a weight, in such a way
that the auxiliary system returns to its initial state at the end of the
process whereas the weight may have risen or fallen.}}

The role of the `weight' in this definition is merely to provide a 
particularly simple source (or sink) of mechanical energy. 
Note that an adiabatic process, physically, does not have to be gentle,
or `static' or anything of the kind. It can be arbitrarily violent! 

\eject

\vglue .75cm

\epsfxsize 16truecm
\epsfysize 7.5truecm 
\epsffile{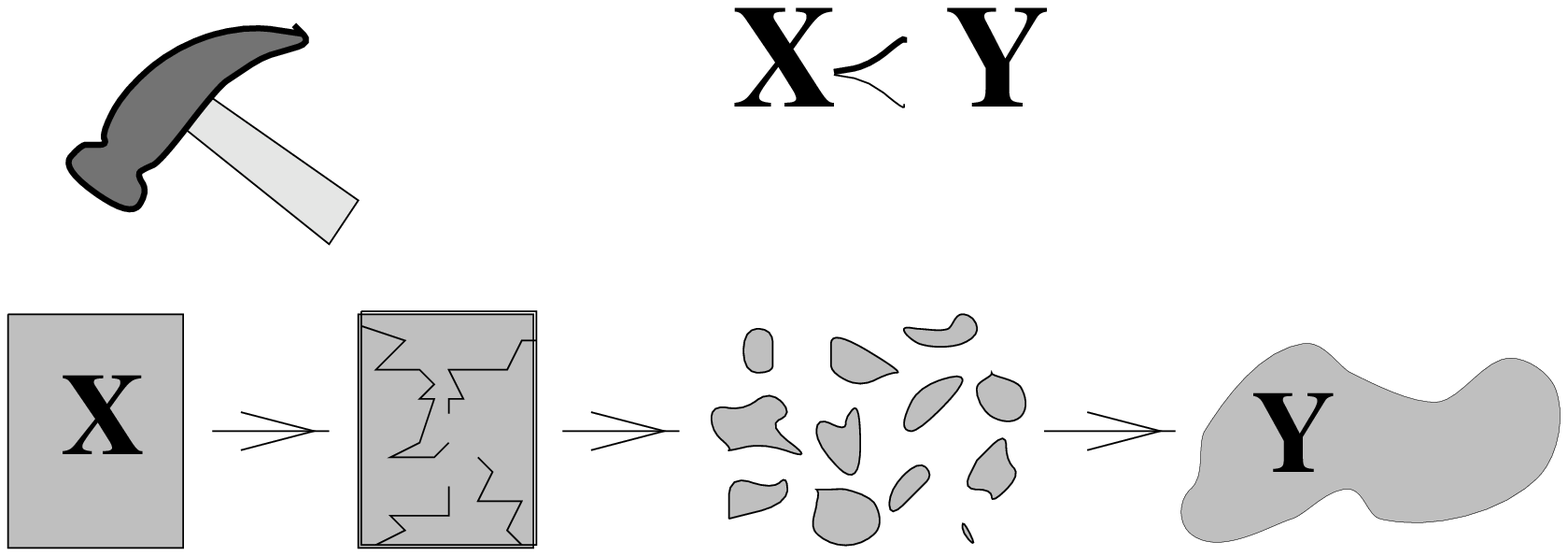}
\indent
{\bf Figure 1.} A violent adiabatic process connecting
equilibrium states $X$ and $Y$.}

\bigskip

An example might be useful here.  Take a pound of hydrogen in a
container with a piston. The states are describable by two numbers,
energy and volume, the latter being determined by the position of the
piston. Starting from some state, $X$, we can take our hand off the
piston and let the volume increase explosively to a larger one.  After
things have calmed  down, call the new equilibrium state $Y$.  Then
$X\prec Y$. Question: Is $Y\prec X$ true? Answer: No. To get from $Y$ to
$X$ we would have to use some machinery and a weight, with the machinery
returning to its initial state, and there is no way this can be done. 
Using a weight we can, indeed, recompress the gas to its original
volume, but we will find that the energy is then larger than 
its original value.   

Let us write
$$
X\prec \prec Y \ \ \  {\rm if}\ \ \  X\prec Y \ \ \ {\rm but\ not}\ \ \
 Y\prec X \ {\rm (written\ }Y\not\prec X). 
$$
In this case we say that we can go from $X$ to $Y$  by an
{\it irreversible adiabatic process}.
If $X\prec Y $ and $Y\prec X$ we say that $X$ and $Y$ are
{\it adiabatically equivalent } and write
\medskip
\centerline{ $X\sima Y\  .$}

Equivalence classes under $\sima$ are called {\it adiabats}.
\smallskip

\item{4.} {\it Comparability:} Given two states $X$ and $Y$ in two
(same or different)  state-spaces, we say that they are comparable if
$X\prec Y$ or $Y\prec X$ (or both). This turns out to be a crucial
notion. Two states are not always comparable; a necessary condition is
that they have the same material composition in terms of the chemical
elements. Example: Since water is ${\rm H}_{2} {\rm O}$ and  the atomic
weights of hydrogen and oxygen are 1 and 16 respectively,  the
states in the compound system of 2 gram of hydrogen and 16 grams of
oxygen are comparable with states in a system consisting of 18 grams of
water (but not with 11 grams of water or 18 grams of oxygen). 

Actually, the classification  of states into various state-spaces is
done  mainly for conceptual convenience.  The second law deals only
with states,  and the only thing we really have to know about any two
of them is whether or not they are comparable.  Given the relation
$\prec$ for all possible states of all possible systems, we can ask
whether this relation can be encoded in an entropy function according to
the following: 

{\bf Entropy principle:} {\it There is a real-valued function on all
states of all systems (including compound systems), called {\bf entropy}
and denoted by $S$ such that

{\parindent=18pt
\item{a)} {{\tt Monotonicity:}} When $X$ and $Y$ are
comparable states then
$$
X\prec Y \hbox{ \ \ {\rm if and only if} \ \ } S(X) \leq S(Y) \ . \eqno(1)
$$
\item{b)} {{\tt Additivity and extensivity:}} If $X$ and $Y$
are states of some (possibly different) systems and if $(X,Y)$ denotes
the corresponding state in the compound system,  then
the entropy is additive for these states, i.e.,
$$
S(X,Y) = S(X) + S(Y) \ .  \eqno (2)
$$
$S$ is also extensive, i.e.,  for or each $\lambda>0$ and each
state $X$ and its scaled copy  $\lambda X \in \Gamma^{(\lambda)}$,
(defined in 2. above) 
$$
S(\lambda X)=\lambda S(X) \ .\eqno(3)
$$
}}

A  formulation logically equivalent to a),   not using 
the word `comparable', is the following pair of statements:
$$
\eqalignno{
X\sima Y &\Longrightarrow S(X) = S(Y) \ \ \ \ {\rm and} \cr
X\prec\prec Y &\Longrightarrow S(X) < S(Y).& (4)\cr }
$$
The last line is especially noteworthy. It says that entropy must
increase in an irreversible  adiabatic  process.

The additivity of entropy in compound  systems is often just taken for
granted, but it is one of the startling conclusions of thermodynamics. 
First of all,  the content of additivity, (2), is considerably more far
reaching than one might think from the simplicity of the notation.
Consider four states $X,X',Y,Y'$ and suppose that $X\prec Y$ and
$X'\prec Y'$.  One of our axioms, A3, will be that then $(X,X')\prec
(Y,Y')$,  and (2) contains nothing new or exciting.  On the other
hand,  the compound system can well have an adiabatic process in which
$(X,X')\prec (Y,Y')$ but $X\not\prec Y$.  In this case, (2) conveys much
information. Indeed, by monotonicity, there will be many cases of this
kind because the inequality $S(X) + S(X') \leq S(Y) + S(Y')$ certainly
does not imply that $S(X) \leq S(Y)$. The fact that the inequality $S(X)
+ S(X') \leq S(Y) + S(Y')$ tells  us {\it exactly } which adiabatic
processes are allowed in the compound system (among  comparable states),
independent of any detailed knowledge of the manner in which the two
systems interact, is astonishing and is at the {\it heart of
thermodynamics.}  The second reason that (2) is startling is this: From
(1) alone, restricted to one system, the function $S$ can be replaced by
$29S$ and still do its job, i.e., satisfy (1). However, (2) says that it
is possible to calibrate  the entropies of all systems (i.e.,
simultaneously adjust all the undetermined multiplicative constants) so
that the entropy $S_{1,2}$ for a compound $\Gamma_1 \times \Gamma_2$ is
$S_{1,2}(X,Y) = S_1(X) + S_2(Y)$, even though systems 1 and 2 are
totally unrelated!

We are now ready to ask some basic questions:
%{\parindent=18pt
\item{Q1:} Which properties of the relation $\prec$ ensure existence and
(essential) uniqueness of $S$?
\item{Q2:} Can these properties be derived from simple physical premises?
\item{Q3:} Which convexity and smoothness properties of $S$ follow from 
the premises?
\item{Q4:} Can temperature 
(and hence an ordering of states by ``hotness" and ``coldness")
 be defined from $S$ and what are its properties?

The answer to question Q1 can be given in the form of six axioms that
are reasonable,  simple, `obvious'  and unexceptionable.  An
additional, crucial assumption is also needed, but we call it a
`hypothesis' instead of an axiom because we show later how it can be
derived from some other axioms, thereby answering question Q2.  

\item{\bf A1.}  {\bf Reflexivity}.  $X \sima X$.
\item{\bf A2.}  {\bf Transitivity}. If $X \prec Y$ and $Y \prec Z$, 
then $X \prec Z$.
\item{\bf A3.} {\bf Consistency}. If $X \prec X^\prime$ and $Y
\prec Y^\prime$, then $(X,Y) \prec
(X^\prime, Y^\prime)$.
\item{\bf A4.} {\bf Scaling Invariance}. If $\lambda > 0$ and
$X \prec Y$,  then
$\lambda X \prec \lambda Y$.
\item{\bf A5.}  {\bf Splitting and Recombination}. 
$X \sima ((1-\lambda) X, \lambda X)$ for all $0 < \lambda <  1$. 
Note that the state-space are not the same on both sides. If $X\in\Gamma$, 
then the state space on the right side is 
$\Gamma^{(1-\lambda)} \times \Gamma^{(\lambda)}$.
\item{\bf A6.}  {\bf Stability}. If
$(X, \varepsilon Z_0) \prec (Y, \varepsilon Z_1)$
for some $Z_0$, $Z_1$ and a sequence of $\varepsilon$'s tending 
to zero, then   $X \prec Y$. This axiom is a substitute for
continuity, which we cannot assume because there is no topology yet. It
says that  `a grain of dust cannot influence the set of adiabatic
processes'. 
%}

An important lemma is that 
(A1)-(A6) imply the {\it cancellation law}, which is used in
many proofs. It says that for any three states $X,Y,Z$ 
$$
 (X,Z)\prec (Y,Z)\ \Longrightarrow \  X\prec Y\ .\eqno(5)
$$

The next concept plays a key role in our treatment. 
\item{\bf CH.}{\bf Definition:} We say that the {\it Comparison
Hypothesis,} (CH), holds for a state-space  $\Gamma$  if  all
pairs  of states in   $\Gamma$ are comparable.

Note that  A3, A4 and A5 automatically extend comparability from a space
$\Gamma$ to certain other cases, e.g., $X\prec ((1-\lambda)Y,\lambda Z)$
for  all $0\leq \lambda \leq 1$  if  $X\prec Y$ and $X\prec Z$.  On
the other hand, comparability on $\Gamma$ alone does not allow us to
conclude that $X$ is comparable to $((1-\lambda)Y,\lambda Z)$ if $X\prec
Y$ but $Z\prec X$. For this, one needs CH on the product space
$\Gamma^{(1-\lambda)}\times \Gamma^{(\lambda)}$, which is not implied by
CH on $\Gamma$.

The significance of A1-A6 and CH is borne out by the following theorem: 

{\bf THEOREM 1 (Equivalence of entropy and A1-A6, given CH).} {\it 
The following are equivalent for a state-space $\Gamma$:

(i) The relation $\prec$ between states in (possibly different) 
multiple scaled copies of $\Gamma$ 
e.g., 
$\Gamma^{(\lambda_1)} \times \Gamma^{(\lambda_2)} \times \cdots\times
\Gamma^{(\lambda_N)}$, 
is
characterized by an entropy function, $S$, on $\Gamma$ in the sense that 
$$
(\lambda_1 X_1,\lambda_2 X_2,\dots)~\prec~(\lambda_1' X_1',\lambda_2'
X_2',\dots)\eqno(6)
$$
is equivalent to  the condition that 
$$\sum_i \lambda_i S(X_i)\leq \sum_j\lambda_j'S(X_j')\eqno(7)$$
whenever 
$$\hbox{$\sum_i$}\lambda_i=\hbox{$\sum_j$}\lambda'_j.\eqno(8)$$

(ii) The relation $\prec$ satisfies conditions (A1)-(A6), and (CH) 
holds for \underbar{every} multiple scaled copy of 
$\Gamma$. 

This entropy function  on $\Gamma$ is unique up to 
affine equivalence, i.e., $S(X)\rightarrow aS(X)+B$, with
$a>0$. }

That (i) $\Longrightarrow$ (ii) is obvious. The proof of (ii)
$\Longrightarrow$ (i) is carried out by an explicit construction of the 
entropy function on $\Gamma$---reminiscent of an old definition of heat by
Laplace and Lavoisier in terms of the amount of ice that a body 
can melt. 

\bigskip

\epsfxsize 15truecm
\epsfysize 7.5truecm 
\epsffile{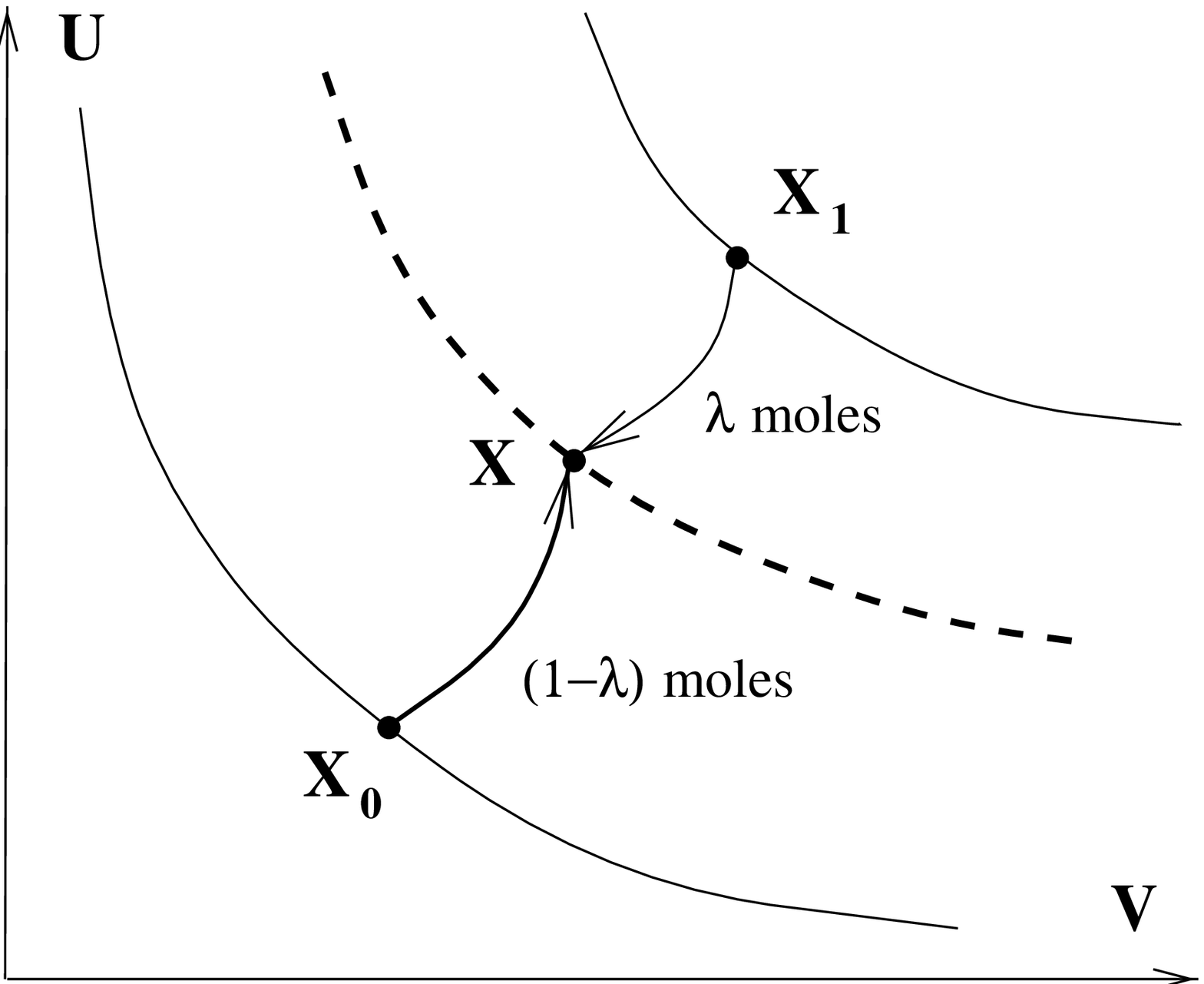}
\indent{\bf Figure 2.} The entropy of $X$ is determined by the 
largest amount of $X_1$ that can be transformed adiabatically into
$X$, with the help of $X_0$.} 
 
\bigskip

{\bf BASIC CONSTRUCTION of $S$:}
Pick two reference points $X_0$ and $X_1$ in $\Gamma$ with
$X_0 \prec\prec X_1$. (If such points do not exist then $S$ is the
constant function.)
Then define for $X\in\Gamma$
$$
S(X):=\sup\{\lambda \, :\, ((1-\lambda)X_0,\lambda X_1)\prec X\, \}.
\eqno(9)
$$

{\it Remarks:} As in axiom A5, two state-spaces are involved in (9).
By axiom A5, $X\sima ((1-\lambda)X, \lambda X)$, and hence, by CH in
the space $\Gamma^{(1-\lambda)}\times \Gamma^{(\lambda)}$, $X$ is
comparable to $((1-\lambda)X_0,\lambda X_1)$.  In (9) we allow $\lambda
\leq  0$ and $\lambda \geq 1$ by using the convention that $(X, -Y)
\prec Z$ means that $X \prec (Y, Z)$ and $(X,0Y)=X$.  For (9) we only
need to know that CH holds in two-fold scaled products of $\Gamma $
with itself. CH will then automatically be true for all products. In
(9)  the reference  points $X_0, X_1$ are fixed and the supremum is
over $\lambda$.  One can ask how $S$ changes if we change the two
points $X_0, X_1$. The answer is that the change is affine, i.e.,
$S(X)\rightarrow aS(X)+B$, with  $a>0$.

Theorem 1 extends to products of multiple scaled
copies of different systems, i.e. to general  {\it compound}  systems.
This extension is an immediate consequence of the following
theorem,  which is proved by applying Theorem 1 to the product of the system
under consideration with some standard reference system.

\eject

{\bf THEOREM 2 (Consistent entropy scales). } {\it Assume that CH
holds for \underbar{all} compound systems.
For each system $\Gamma$ let
$S_{\Gamma}$ be some definite entropy function on $\Gamma$ in the sense
of Theorem 1. Then
there are constants $a_{\Gamma}$ and $B{(\Gamma)}$ such that the
function $S$, defined for all states of all systems by
$$
S(X)= a_{\Gamma} S_{\Gamma} (X)+ B{(\Gamma)}\eqno(10)
$$
for $X\in \Gamma$, 
satisfies additivity (2), extensivity (3), and monotonicity (1) in the 
sense that whenever 
$X$ and $Y$ are  in the same state space then
$$
X\prec Y \quad\quad \hbox{\rm if and only if} \quad\quad S(X)\leq
S(Y).\eqno(11)
$$ 

}

Theorem 2 is what we need, except for the 
question of mixing and chemical reactions, which is treated at the end 
and which can be put aside at a first reading.  In other words, as 
long as we do not consider adiabatic processes in which systems are 
converted into each other (e.g., a compound system consisting of 
vessel of hydrogen and a vessel of oxygen is converted into a vessel 
of water), the entropy principle has been verified.  If that is so, 
what remains to be done, the reader may justifiably ask?  The answer 
is twofold: First, Theorem 2 requires that CH holds for {\it all }
systems, and we are not content to  take  this as an axiom.  Second,
important notions of thermodynamics  such as
`thermal equilibrium'  (which will eventually 
lead to a precise definition of `temperature' ) have not 
appeared so far. We shall see that these two points (i.e., thermal
equilibrium and CH) are not unrelated. 

As for CH, other
authors, [6], [2], [4] and [9] essentially {\it postulate} 
that it holds for all systems by making
it axiomatic that comparable states fall into equivalence classes. (This
means that the conditions $X\prec Z$ and $Y\prec Z$ always imply that
$X$ and $Y$ are comparable: likewise, they must be comparable if $Z\prec
X$ and $Z\prec Y$).   By identifying a `state-space' with an
equivalence class, the comparison hypothesis then holds in these other
approaches  {\it by  assumption}  for all state-spaces.   We, in
contrast, would like to derive CH from something that we consider more
basic.  Two ingredients will be needed: The analysis of certain special,
but commonplace systems called `simple systems' and some assumptions
about thermal contact (the `Zeroth law') that will act as a kind of glue
holding the parts of a compound systems in harmony with each other.

A {\bf Simple System} is one whose state-space can be identified with
some open  convex subset of some ${\bf R}^{n+1}$ with a distinguished
coordinate denoted by $U$, called the {\it energy}, and additional
coordinates $V\in {\bf R}^{n}$, called  {\it work coordinates.}  The
energy coordinate is the way in which thermodynamics makes contact with
mechanics, where the concept of energy arises and is precisely defined. 
The fact that the amount of energy in a state is independent of the
manner in which the state was arrived at is, in reality, the  first
law of thermodynamics.  A typical (and often the only) work coordinate
is the volume of a fluid or gas (controlled by a piston); other examples
are deformation coordinates of a solid or magnetization of a
paramagnetic substance.  

Our goal is to show, with the addition of a few more axioms, that 
CH holds for simple systems and their scaled products. In the process, 
we will introduce more structure, which will capture the intuitive notions 
of thermodynamics; thermal equilibrium is one.

First, there is an axiom about convexity: 

{\parindent =18pt
\item{{\bf A7.}} {\bf Convex combination.} If $X$ and $Y$ are states of
a simple system and $t \in [0,1]$ then 
$$(t X, (1-t) Y) ~\prec ~t X +
(1-t)Y\ ,
$$ 
in the sense of ordinary convex addition of points in ${\bf R}^{n+1}$. 
A straightforward consequence of this axiom (and A5)  is that the 
{\bf forward  sectors}
$$
A_{X}:=\{Y\in\Gamma :X\prec Y\}\eqno(12)
$$ 
{\hskip18pt}of states $X$ in a simple system $\Gamma$ are {\it convex} 
sets. 
}

\vglue .75cm

\epsfxsize 15truecm
\epsfysize 7.5truecm 
\epsffile{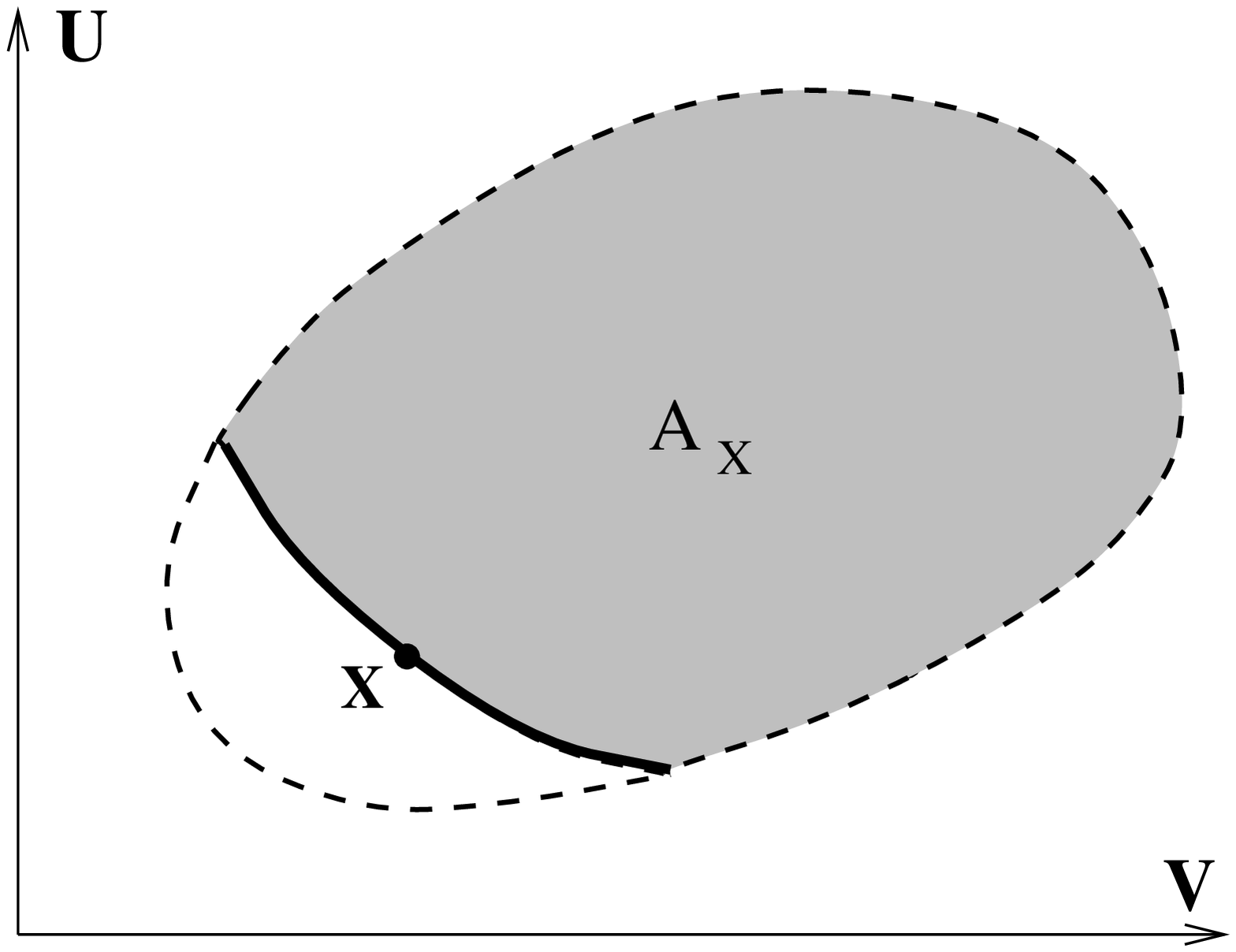}
\indent
{\bf Figure 3.} The coordinates $U$ and $V$ of a simple system.
The state space (bounded by dashed line)  and the forward sector $A_X$
(shaded) of  a state  $X$ are convex, by axiom A7. The boundary of
$A_X$ (full line) is an adiabat.

\eject

Another consequence is a connection between the existence of 
irreversible processes and  Carath\'eodory's  principle  ([3], [1]) mentioned
above.

{\bf LEMMA 1.} {\it Assume (A1)-(A7) for
$\Gamma\subset R^N$ and consider the following statements:

(a). {\rm Existence of irreversible processes:} For every $X\in\Gamma$
there is a $Y\in\Gamma$ with $X\prec\prec Y$.

(b). {\rm Carath\'eodory's principle:} In every neighborhood of every
$X\in\Gamma$
there is a $Z\in\Gamma$ with $X\not\prec Z$.

Then (a) $\Longrightarrow$(b) always. If the forward sectors in $\Gamma$
have interior points, then (b) $\Longrightarrow$ (a).}

We need three more axioms for simple systems, which will take 
us into an analytic detour. The first of these
establishes (a) above.

{\parindent =18pt
\item{{\bf A8.}} {\bf Irreversibility.} For each $X \in \Gamma$ there
is a point $Y \in \Gamma$ such that $X \prec\prec Y$. (This axiom
is implied by A14 below, but is stated here separately because 
important conclusions can be drawn from it alone.)
\item{{\bf A9.}} {\bf Lipschitz tangent planes.} For each $X\in \Gamma$
the {\it forward sector} $A_X=\{Y\in\Gamma:X\prec Y\}$ has a {\it unique}
support
plane at $X$ (i.e.,
$ A_X$ has a {\it tangent plane} at $X$).
The tangent plane is
assumed to be a {\it locally Lipschitz continuous} function of $X$, 
in the sense explained below.
\item{{\bf A10.}} {\bf Connectedness of the boundary.} The boundary
$\partial A_X$ (relative to the open set $\Gamma$) of every forward
sector $A_{X}\subset\Gamma$ is connected. (This is technical and
conceivably can be replaced by something else.)

}

 %Note that every forward sector $A_{X}$ is convex, by axiom A7. 
Axiom A8  plus
Lemma 1 asserts that every $X$ lies on the boundary $\partial A_X$ of 
its forward sector. Although axiom A9 asserts that
the convex set, $A_X$, has a true tangent at $X$ only, it is an easy
consequence of axiom A2 that $A_X$ has a true tangent everywhere on its
boundary. To say that this tangent plane is locally Lipschitz 
continuous means that if $X =
(U^0, V^0)$ then this plane is given by
$$
U - U^0 + \sum\nolimits^n_1 P_i (X) (V_i - V^0_i) ~=~ 0.\eqno(13)
$$
with locally Lipschitz continuous functions $P_i$. 
The function $P_i$ is called the generalized {\it pressure} conjugate to
the work coordinate $V_i$ .  (When $V_i$ is the volume, $P_i$ is the
ordinary pressure.)

Lipschitz continuity and connectedness is a well known guarantee that the
coupled differential equations 
$$
{{\partial} U \over {\partial}V_j} (V) ~= ~- P_j (U (V), V)
\quad {\rm for} \ j=1,\dots ,n \ . \eqno(14)
$$
not only have a solution (since we know that the surface $\partial A_X$
exists) but this solution must be unique. Thus, if
$Y\in \partial A_X$ then $X\in \partial A_Y$. In short, the 
surfaces $\partial A_X$ foliate the state-space $\Gamma$.
What is less obvious, but very important because it instantly
gives us the comparison hypothesis for $\Gamma $, is the following.

{\bf THEOREM 3 (Forward sectors are nested).}  {\it  If $A_X$
and $A_Y$ are two forward sectors in the state-space, $\Gamma$, of a
simple system then exactly one of the following holds.

(a).  $A_X = A_Y$, i.e., $ X\sima Y$.

(b).  $A_X \subset {\rm Interior}(A_Y)$, i.e., $Y \prec\prec X$
%\medskip

(c).  $A_Y \subset {\rm Interior}(A_X)$, i.e., $X \prec\prec Y$.
}
\medskip
\epsfxsize 12.5cm
\epsfysize 14.5truecm 
\epsffile{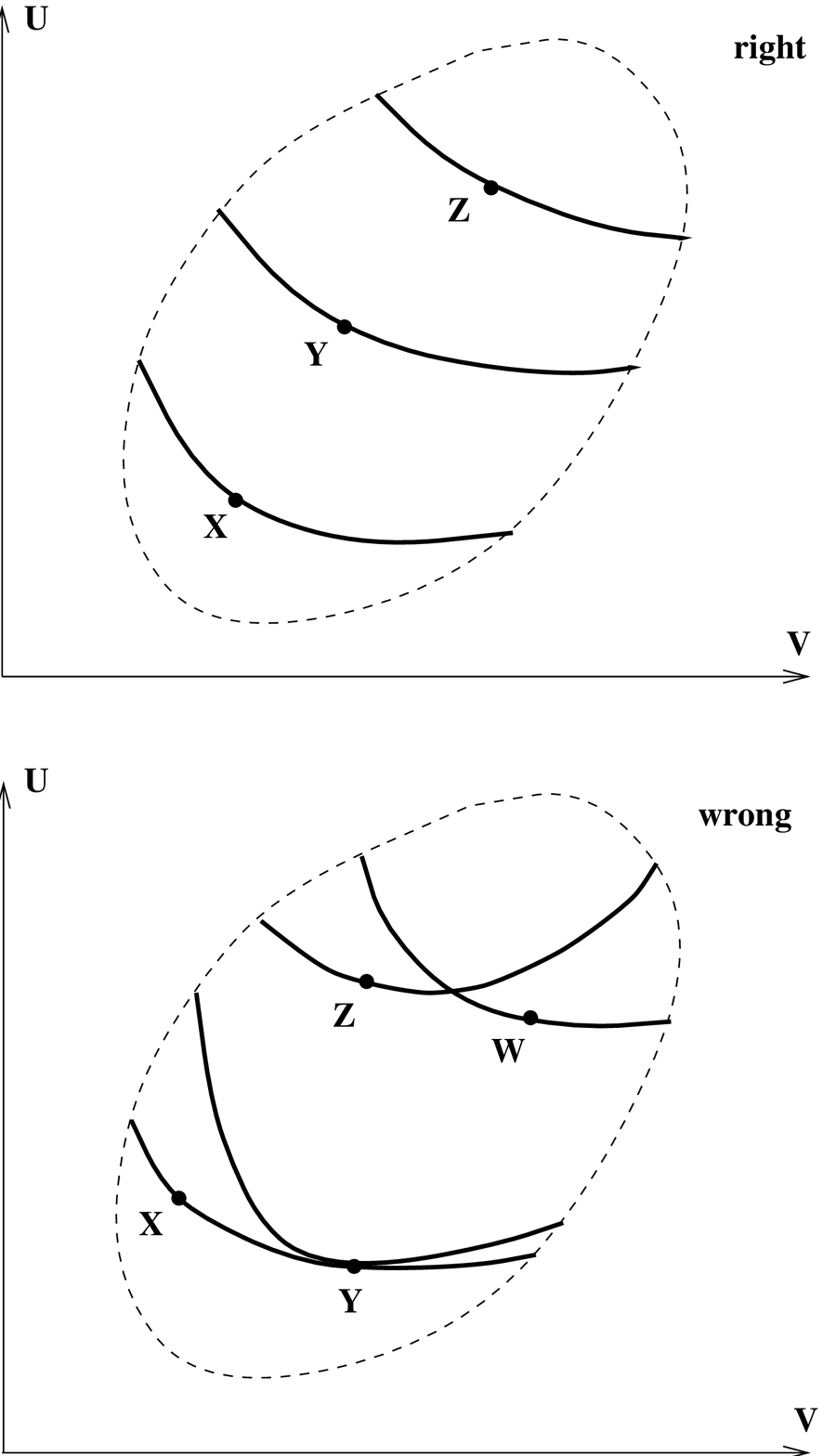}
\indent
{\bf Figure 4.} The forward sectors of a simple system are
nested.  The bottom figure shows what could, in principle, could go 
wrong---but does not.

It can also be shown from our axioms that the orientation of forward
sectors w.r.t. the energy axis is the same for {\it all} simple systems.
By convention we choose the direction  of the energy axis so that the
the energy  always  {\it increases} in adiabatic processes at fixed
work coordinates. When temperature is defined later, this will imply
that temperature is always positive.

Theorem 3 implies that $Y$ is on the boundary of $A_X$ if and only if
$X$ is on the boundary of $A_Y$. Thus the  adiabats, i.e., the $\sima$
equivalence classes, consist of these boundaries.

Before leaving the subject of simple systems let us remark on the
connection with Carath\'eo\-dory's development. The point of  contact is
the fact that $X\in \partial A_X$. We assume that $A_X$ is convex and
use transitivity and Lipschitz continuity to arrive, eventually, at
Theorem 3. Carath\'eo\-dory uses Frobenius's theorem, plus assumptions about
differentiability to conclude the existence---locally---of a surface
containing $X$.  Important {\it global} information, such as Theorem 3,
are then not easy to obtain without further assumptions,  as discussed, 
e.g., in [1].  

The next topic is {\it thermal contact}  and the zeroth law,  which
entails the very special assumptions about $\prec$ that we mentioned 
earlier.  It will enable us to establish CH for products of several
systems, and thereby show, via Theorem 2, that entropy exists and is
additive. Although we have established CH for a simple system, $\Gamma$,
we have not yet established CH even for a product of two copies of
$\Gamma$. This is needed in the definition of $S$ given in (9).  The
$S$ in (9)  is determined up to an affine shift and we want to be able
to calibrate the entropies (i.e., adjust the multiplicative and additive
constants) of all systems so that they work together to form a global
$S$ satisfying the entropy principle.  We need five more axioms.  They
might look a bit abstract, so a few words of introduction might be
helpful.

In order to relate systems to each other, in the hope of establishing CH
for compounds, and thereby an additive entropy function, some way must
be found to put them into contact with each other. Heuristically, we
imagine two simple systems (the same or different) side by side, and fix
the work coordinates (e.g., the volume) of each. Connect them with a
``copper thread'' and wait for equilibrium to be established. The total
energy $U$ will not change but the individual energies, $U_1$ and $U_2$
will adjust to  values that depend on $U$ and the work coordinates. 
This new system (with the thread permanently connected) then behaves
like a simple system (with one energy coordinate) but with several work
coordinates (the union of the two work coordinates). 
Thus, if we start
initially with 
$X_1=(U_1, V_1)$ for system 1 and 
$X_2=(U_2, V_2)$  for
system 2, and if we end up with $X=(U, V_1, V_2)$ for the new system, we
can say that $(X_1, X_2) \prec X $. This holds for every choice of  $U_1
$ and $U_2$ whose sum is $U$. Moreover, after thermal equilibrium is
reached, the two systems can be disconnected, if we wish, and once more
form a compound system, whose component parts we say are in thermal
equilibrium. That this is transitive is the zeroth law.  

Thus, we can not only make compound systems consisting of independent
subsystems (which can interact, but separate again), we can also make a
new simple system out of two simple systems.  To do this an energy
coordinate has to disappear, and thermal contact does this for us. All
of this is formalized in the following three axioms.

%which we state and  then comment on later.

{\parindent=18pt
\item{\bf A11.} {\bf Thermal contact.} For any two simple systems with 
state-spaces $\Gamma_1$ and $\Gamma_2$, there is another {\it simple}
 system, called the {\it thermal join}
of $\Gamma_1$ and $\Gamma_2$,  with
state-space
$$
\Delta_{12} = \{ (U,V_1,V_2) : U=U_1+U_2 \;{\rm with}\; 
(U_1,V_1)\in \Gamma_1,
(U_2,V_2)\in \Gamma_2\}.\eqno(15)
$$
Moreover,
$$
\Gamma_1\times \Gamma_2 \ni ((U_1,V_1), \ (U_2,V_2)) \prec
(U_1+U_2, V_1,V_2) \in \Delta_{12}.\eqno(16)
$$
\item{\bf A12. } {\bf Thermal splitting.}
For any point $(U,V_1,V_2) \in \Delta_{12}$ there is at least
one pair of
states, $(U_1,V_1) \in \Gamma_1$, $(U_2,V_2))\in
\Gamma_2$, with $U=U_1+U_2$,  such that
$$
(U,V_1,V_2)\sima ((U_1,V_1), (U_2,V_2)).\eqno(17)
$$
If $(U,V_1,V_2)\sima ((U_1,V_1), (U_2,V_2))$ we say that 
the states $X=(U_1,V_1)$ and $Y= (U_2,V_2))$ are in {\it thermal 
equilibrium } and write 
$$
X\simt Y.
$$
\item{\bf A13.} {\bf Zeroth law of thermodynamics.} If $X\simt Y$ and if
$Y\simt Z$ then $X\simt Z$.  

}

%A11 says that one can make two simple systems into one by an adiabatic
%process, which requires removing one energy coordinate, of course (but
%not work coordinates).  This
%is done by fiat.  
 
A11 and A12 together say that for each choice of the individual work
coordinates there is a way to divide up the energy $U$ between  the two
systems in a stable manner. A12 is the stability statement, for it says
that joining is reversible, i.e., once the equilibrium has been
established, one can  cut the copper thread and retrieve the two systems
back again, but with  a special partition of the energies.  

This reversibility allows us to think of the thermal join, which is a
simple system in its own right, as a special subset of the product
system, $\Gamma_1 \times \Gamma_2$, which we call the 
{\it thermal diagonal}. In particular, A12 allows us to prove easily that
$X\simt \lambda X$ for all $X$ and all $\lambda >0$.

A13 is the famous zeroth law, which says that the thermal equilibrium is
transitive, and hence an equivalence relation. Often this law is taken
to mean that there the equivalence classes can be labeled by an
`empirical' temperature, but we do not want to mention temperature at
all at this point. It will appear later.

Two more axioms are needed. 

A14  requires that for every adiabat (i.e., an equivalence class w.r.t.\
$\sima$\ ) there exists at least one isotherm (i.e., an equivalence
class w.r.t.  $\simt$\ ), containing points on both sides of the
adiabat.  Note that, for each given $X$, only two points in the entire
state space $\Gamma$ are required to have the stated property. This
assumption essentially prevents a state-space from breaking up into two
pieces that do not  communicate with each other.  Without it,
counterexamples to CH for compound systems can be constructed.
A14 implies A8,  but we listed
A8 separately in order not to  confuse the discussion of simple
systems with thermal equilibrium.

\medskip

\epsfxsize 16.5truecm
\epsfysize 8.5truecm 
\epsffile{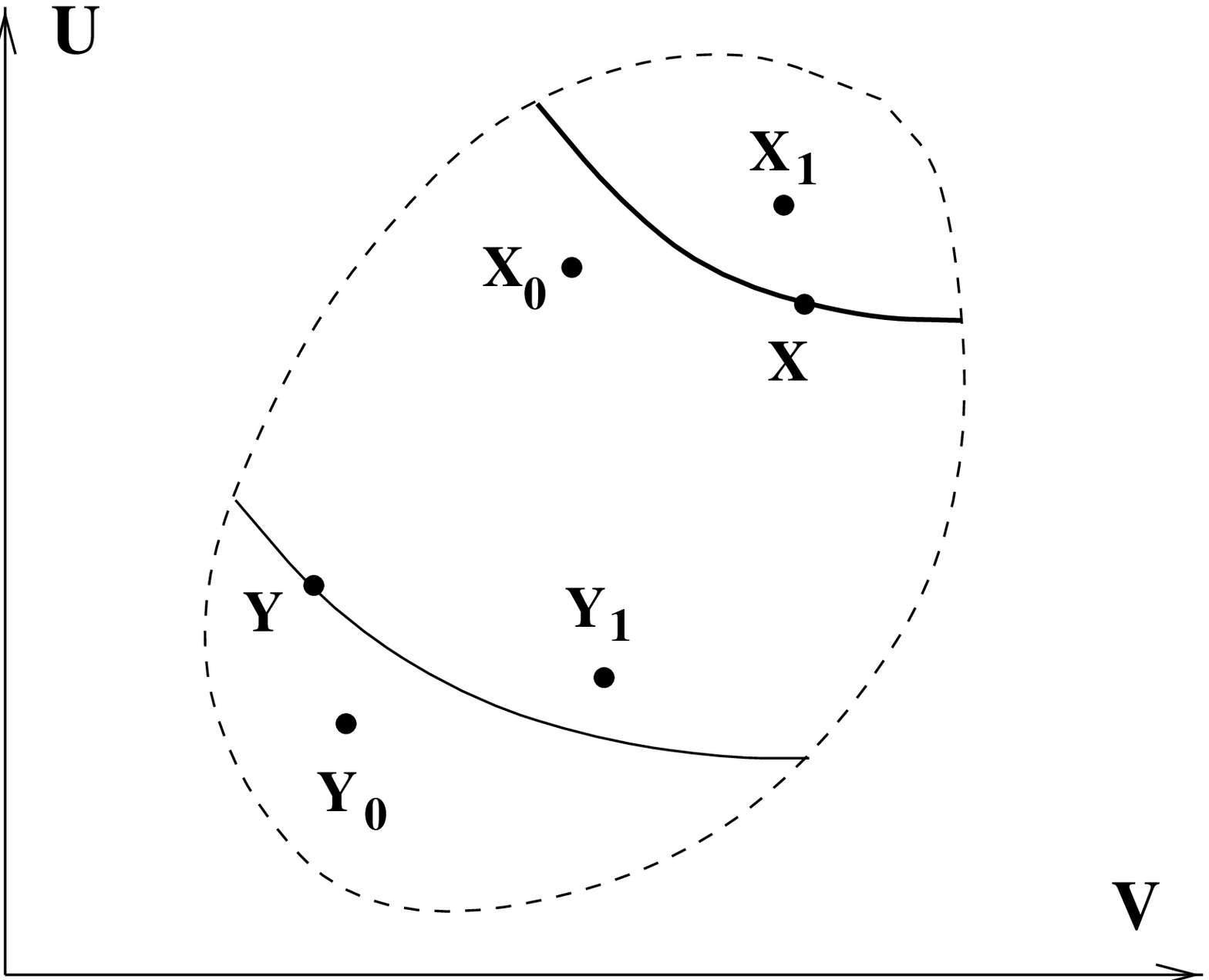}
\indent
{\bf Figure 5.} Transversality, A14, requires that each $X$ have
points on each side of its adiabat that are in thermal equilibrium.

\bigskip

A15 is a technical and perhaps can be eliminated. Its physical
motivation is that a sufficiently large copy of a system can act as a
heat bath for other systems. When temperature is introduced later, A15
will have  the meaning that all systems have the same temperature range. 
This postulate is needed if we want to be able to bring every system
into thermal equilibrium with every other system.

\item {\bf A14.} {\bf Transversality.} If
$\Gamma$ is the state space of a simple system and if $X \in \Gamma$,
then there exist states $X_0\simt X_1$ with $X_0\prec\prec X\prec\prec
X_1$.
\item{\bf A15.} {\bf Universal temperature range.} If $\Gamma_1$ and
$\Gamma_2$ are state spaces of simple systems then,  for every
$X\in\Gamma_1$ and every $V$ belonging to the projection of
$\Gamma_2$ onto the space of its work coordinates,  there is a
$Y\in\Gamma_2$ with work coordinates $V$ such that $X\simt Y$.

The reader should note that the concept  `thermal contact' has
appeared, but not temperature or hot and cold or anything resembling the
Clausius or Kelvin-Planck formulations of the second law.  Nevertheless,
we come to the main achievement of our approach: {\it With these axioms
we can establish CH for products of simple systems} (each of which
satisfies CH, as we already know).   First, the thermal join
establishes CH for  the (scaled) product of a simple system with itself.
The basic idea here is that the points in the product that lie on the
thermal diagonal  are comparable, since points in a simple system are
comparable. In particular, with $X, X_{0}, X_{1}$ as in A14, the states
$((1-\lambda)X_{0}, \lambda X_{1})$ and $((1-\lambda)X, \lambda X)$ can
be regarded as states of the {\it same} simple system and are,
therefore, comparable.  {\it This is the key point needed for the
construction of $S$, according to (9).}  The importance of
transversality is thus brought into focus.

With some more work we can establish CH for multiple scaled copies of a
simple system. Thus, we have established $S$ within the context of one
system and copies of the system, i.e. condition (ii) of Theorem 
1. As long as we stay within such a group
of systems there is no way to determine the unknown multiplicative or
additive entropy constants. The next task is to show that the
multiplicative constants can be adjusted to give a universal entropy
valid for copies of {\it different} systems, i.e. to establish the 
hypothesis of Theorem 2.  This is based on the
following.

{\bf LEMMA 4 (Existence of calibrators).} {\it If $\Gamma_1$ and
$\Gamma_2$ are simple systems, then there exist states
$X_0,X_1\in\Gamma_1$ and $Y_0,Y_1\in\Gamma_2$ such that
$$
X_0\prec\prec X_1 \quad\quad {\rm and}\quad\quad Y_0\prec\prec Y_1
$$
and
$$
(X_0,Y_1)\sima (X_1,Y_0).
$$
}

The significance of Lemma 4 is that it allows us to fix  the 
{\it multiplicative} constants by the condition
$$
S_1(X_0) + S_2(Y_1) = S_1(X_1)+S_2(Y_0).\eqno(18)
$$

The proof of Lemma 4 is complicated and really uses  all the axioms A1
to A14. With  its aid we arrive at our chief goal, which is CH for 
compound systems. 

{\bf THEOREM 4 (Entropy principle in products of simple systems).} {\it
The comparison hypothesis CH is valid in arbitrary scaled products of
simple systems. Hence, by Theorem 2, the relation $\prec$ among states
in such state-spaces is characterized by an entropy function $S$. The
entropy function is \underbar {unique}, up to an overall multiplicative
constant and one additive constant for each simple system under
consideration.}

At last, we are now ready to  define  {\it temperature}.  Concavity of
$S$  (implied by A7),  Lipschitz
continuity of the pressure and the transversality condition, together
with some real analysis, play key roles in the following, which answers
questions Q3 and Q4 posed at the beginning.

{\bf THEOREM 5 (Entropy defines temperature.)}
{\it The entropy, $S$,  is a concave and continuously 
differentiable function on the
state space of a simple system. If the function $T$ is defined by
$$
{1 \over T}~:=\left({\partial S\over \partial U}\right)_V\eqno(19)
$$
then $T > 0 $  and $T$ characterizes the relation $\simt$ in the
sense that $X\simt Y$ if and only if $T(X)=T(Y)$.  Moreover, if two
systems are brought into thermal contact with fixed work coordinates
then, since the total entropy cannot decrease, the  energy flows from
the system with the higher $T$ to the system with the lower $T$.}

The temperature need not be a strictly monotone function of $U$; indeed,
it is not so in a `multiphase region'. It follows  that $T$ is not
always  capable of specifying  a state, and this fact  can cause some
pain in traditional discussions of the second law---if it is recognized,
which usually it is not.  
\medskip

{\bf  Mixing and Chemical Reactions.} The core results of our 
analysis have now been presented and readers satisfied with the 
entropy principle in the form of Theorem 4 may wish to stop
at this point.  Nevertheless, a
nagging doubt will occur to some, because there are important adiabatic
processes in which systems are not conserved, and these processes are
not yet covered in the theory. A critical 
study of the usual textbook treatments should
convince the reader that this subject is not easy, but
in view of the manifold applications of thermodynamics to chemistry and 
biology it is important to tell the whole story
and not  ignore such processes.

One can formulate the problem as the determination of 
the additive constants
$B{(\Gamma)}$ of Theorem 2. As long as we consider only adiabatic
processes that preserve the amount  of each simple system (i.e., such
that Eqs. (6) and (8) hold), these constants are indeterminate. This is
no longer the case, however, if we consider mixing processes and
chemical reactions (which are not really different, as far as
thermodynamics is concerned.) It then becomes a nontrivial question
whether the additive constants can be chosen in such a way that the
entropy principle holds. Oddly, this determination turns out to be far
more complex, mathematically and physically  than the determination of
the multiplicative constants (Theorem 2). In traditional treatments one
usually resorts to {\it gedanken} experiments involving strange, nonexistent
objects called `semipermeable' membranes and `van t'Hofft boxes'.  We
present here a general and rigorous approach which avoids all this.

What we already know is that every system has a 
well-defined entropy function, e.g., for each  $\Gamma$ 
there is $S_\Gamma$, and we know from Theorem 2 that the multiplicative 
constants $a_{\Gamma}$ can been
determined
in such a way that the sum of the entropies increases in any  adiabatic
process in any compound space
$\Gamma_1 \times  \Gamma_2 \times ...$. Thus, if
$X_i \in  \Gamma_i$ and $Y_i \in  \Gamma_i$ then
$$
(X_1,X_2,...) \prec (Y_1,Y_2,...) \quad {\rm if \ and \ only \ if} \quad
\hbox{$\sum_{i}$}S_i(X_i)\leq \hbox{$\sum_{j}$}S_j(Y_j). \eqno(20)$$
where we have denoted $S_{\Gamma_i}$ by $S_i$ for short.  The additive
entropy constants do not matter here since each function $S_i$ appears
on both sides of this inequality. It is important to note that this
applies even to processes  that, in intermediate steps,  take one system
into another, provided the total compound system is the same at the
beginning and at the end of the process.  

The task is to find 
constants $B(\Gamma)$, one for each 
state space $\Gamma$, in such a way that the 
entropy defined by
$$
S(X) := S_\Gamma(X) + B(\Gamma)  \quad\quad {\rm for} \quad\quad X \in 
\Gamma    \eqno(21)
$$
satisfies
$$
S(X) \leq S(Y)   \eqno(22)
$$
whenever 
$$
X\prec Y  \quad\quad {\rm with } \quad\quad X \in 
\Gamma, \ Y \in \Gamma'. 
$$
Additionally, we require that the  newly defined entropy satisfies
scaling and additivity under composition. Since the initial entropies
$S_\Gamma(X)$ already satisfy them, these requirements become 
conditions on the additive constants $B(\Gamma)$:
$$
B(\Gamma_1^{(\lambda_1)}\times 
\Gamma_2^{(\lambda_2)})~=~\lambda_1B(\Gamma_1)+\lambda_2 B(\Gamma_2)\eqno(23)
$$
for all state spaces $\Gamma_1$, $\Gamma_2$ under considerations and 
$\lambda_1,\lambda_2>0$.
Some reflection shows us that consistency in the definition of the
entropy constants $B(\Gamma)$ requires us to consider all possible
chains of adiabatic processes leading from one space to another via
intermediate steps.  Moreover, the additivity requirement leads us to
allow the use of a `catalyst' in these processes, i.e., an auxiliary
system, that is recovered at the end, although a state change {\it
within} this system might take place.  With this in mind we define
quantities $F(\Gamma,\Gamma')$ that incorporate the entropy differences
in all such chains leading from $\Gamma$ to $\Gamma'$.  These are built
up from simpler quantities $D(\Gamma,\Gamma')$, which measure the entropy
differences in one-step processes, and $E(\Gamma,\Gamma')$, where the
`catalyst' is absent. The precise definitions are as follows. First, 
$$
D(\Gamma,\Gamma') := ~\inf \{S_{\Gamma'}(Y)-S_{\Gamma}(X) \ : \ 
X\in\Gamma,\ Y\in\Gamma' ,\ X \prec Y \}.             \eqno(24)
$$
If there is no adiabatic process leading from $\Gamma$ to $\Gamma'$ we put 
$D(\Gamma,\Gamma')=\infty$. Next, 
for any
given $\Gamma$ and $\Gamma'$ we consider all 
finite chains of state spaces, 
$\Gamma=\Gamma_1,\Gamma_2,\dots,\Gamma_N=\Gamma'$
such that $D(\Gamma_i,\Gamma_{i+1})<\infty$ for all i, and we define
$$
E(\Gamma,\Gamma'):=~\inf\{D(\Gamma_1,\Gamma_{2})+
\cdots +D(\Gamma_{N-1},\Gamma_{N})
\}, 
\eqno(25)
$$
where the infimum is taken over all such chains linking $\Gamma$ with 
$\Gamma'$. Finally
we define
$$
F(\Gamma,\Gamma'):=~\inf\{E(\Gamma\times\Gamma_0, \Gamma'\times
\Gamma_0)\} \ , 
\eqno(26)
$$
where the infimum is taken over all state spaces $\Gamma_0$. (These 
are the `catalysts'.)

The importance of the $F$'s for the determination of the additive
constants is made clear in the following theorem: 
\medskip

{\bf THEOREM 6 (Constant entropy differences).} {\it
If $\Gamma$ and  $\Gamma'$ are two state spaces 
then for any two states $X\in \Gamma$ and  $ Y\in \Gamma'$ 
$$
X\prec Y \quad \hbox{\rm if and only if} \quad S_\Gamma(X) +F(\Gamma,
\Gamma') ~\leq ~S_{\Gamma'}(Y) . \eqno(27)
$$}
An essential ingredient for the proof of this theorem is Eq.\ (20). 

According to Theorem 6 the  determination of the entropy 
constants $B(\Gamma)$ amounts to satisfying the inequalities
$$
-F(\Gamma',\Gamma)~\leq ~B(\Gamma)-B(\Gamma')~\leq~ F(\Gamma,\Gamma')
\eqno(28)
$$
together with the linearity condition (23). It is clear that (28) can
only be satisfied with finite constants $B(\Gamma)$ and $B(\Gamma')$, if
$F(\Gamma,\Gamma')>-\infty$. To exclude the pathological case 
$F(\Gamma,\Gamma')=-\infty$ we
introduce our last  axiom A16, whose statement requires  the following
definition. 

{\bf Definition.} A state-space, $\Gamma$ is said to be {\it connected} to
another state-space $\Gamma'$  if there are states $X\in \Gamma$ and
$Y\in\Gamma'$, and state spaces $\Gamma_1,\dots,\Gamma_N$ with states
$X_i, Y_i\in\Gamma_i$, $i=1,\dots,N$, and a state space $\Gamma_0$ with
states $X_0,Y_0\in\Gamma_0$, such that 
$$
(X,X_0)\prec Y_1,\quad~~ X_i\prec  Y_{i+1},~i=1,\dots, N-1,\quad~~
X_N\prec (Y,Y_0).
$$

{\parindent=18pt 
\item{\bf A16.} {\bf (Absence of sinks)}: If $\Gamma$
is connected to $\Gamma'$ then $\Gamma'$ is connected to $\Gamma$.

}
This axiom excludes $F(\Gamma,\Gamma')=-\infty$ because, on
general grounds,  one   always has 
$$
-F(\Gamma',\Gamma)~\leq ~F(\Gamma,\Gamma') \ .      \eqno(29)
$$
Hence $F(\Gamma,\Gamma')=-\infty$ (which means, in particular, that
$\Gamma$ is connected to $\Gamma'$) would imply
$F(\Gamma',\Gamma)=\infty$, i.e., that there is no way back from
$\Gamma'$ to $\Gamma$. This is excluded by Axiom 16.

The quantities $F(\Gamma,\Gamma')$ have simple subadditivity properties
that allow us to use the Hahn-Banach theorem to satisfy the inequalities
(28), with constants $B(\Gamma)$ that depend linearly on $\Gamma$, in
the sense of Eq.\ (23). Hence we arrive at

{\bf THEOREM 7 (Universal entropy).} {\it The additive entropy constants
of all systems can be calibrated in such a way that the entropy is
additive and extensive, and $X\prec Y$ implies $S(X)\leq S(Y)$, even
when $X$ and $Y$ do not belong to the same state space.}
\medskip
Our final remark concerns the remaining non-uniqueness of the constants 
$B(\Gamma)$.
This indeterminacy  can be traced back to the non-uniqueness of
a linear functional lying between $-F(\Gamma',\Gamma)$ and
$F(\Gamma,\Gamma')$ and has two
possible sources: One is that some pairs of state-spaces $\Gamma$ and
$\Gamma'$ may not be connected, i.e., $F(\Gamma,\Gamma')$ may be infinite
(in which case $F(\Gamma',\Gamma)$ is also infinite by axiom A16). 
The other is that there might be a true gap, i.e.,
$$
-F(\Gamma',\Gamma)~<~F(\Gamma,\Gamma')\eqno(32)
$$
might hold for some state spaces, even if both sides are finite.

In nature only states containing the same amount of the chemical
elements can be transformed into each other. Hence
$F(\Gamma,\Gamma')=+\infty$ for many pairs of state spaces, in
particular, for those that contain different amounts of some chemical
element.  The constants $B(\Gamma)$ are, therefore, never unique: For
each equivalence class of state spaces (with respect to the relation of
connectedness) one can define a constant that is arbitrary except for
the proviso that the constants should be additive and extensive under
composition and scaling of systems. In our world there are  92 chemical
elements (or, strictly speaking, a somewhat larger number, $N$, since
one should count different isotopes as different elements), and this
leaves us with at least 92 free constants that specify the entropy of
one gram of each of the chemical elements in some specific state.  

The other possible source of non-uniqueness, a nontrivial gap (32) for 
systems 
with the same composition in terms of the chemical elements is, as
far as we know, not realized in nature. (Note that this assertion can be 
tested 
experimentally without invoking semipermeable membranes.)  Hence, once the 
entropy constants for the chemical
elements have been fixed and a temperature unit has been chosen (to fix the 
multiplicative constants) the universal entropy is completely fixed.

%\smallskip
We are indebted to many people for helpful discussions, including Fred
Almgren, Thor Bak, Bernard Baumgartner, Pierluigi Contucci, Roy
Jackson, Anthony Knapp, Martin Kruskal, Mary Beth Ruskai and Jan Philip
Solovej.

\bigskip %%%%%%%%%%%% 
\centerline{{\bf REFERENCES} } 
\parindent=0pt

\item{[1]} Boyling, J.B., 1972 {\it An axiomatic approach to classical
thermodynamics,} Proc.~Roy.~Soc.\ Lon\-don {\bf A329}, 35-70.

\item{[2]} Buchdahl, H.~A., 1966, {\it The Concepts of Classical
Thermodynamics,} (Cambridge University Press, Cambridge).  

\item{[3]} Carath\'eodory, C., 1909 {\it  Untersuchung \"uber die
Grundlagen der Thermodynamik,} Math.\ Ann\-alen {\bf 67}, 355-386.

\item{[4]} Cooper, J.L.B., 1967 {\it The foundations of thermodynamics,}
Jour.~Math.~Anal.~and Appl. {\bf 17}, 172-193.  

\item{[5]} Duistermaat, J.~J., 1968, {\it Energy and entropy as  real
morphisms for addition and order,} Synthese {\bf 18}, 327-393.

\item{[6]} Giles, R., 1964, {\it Mathematical Foundations of
Thermodynamics,} (Pergamon, Oxford).

\item{[7]} Lieb, E.H. and Yngvason, J., {\it The Physics and Mathematics
of the Second Law of Thermodynamics,} Preprint (1997). Physics Reports
(in press). Austin Math. Phys. archive 97-457. Los Alamos archive
cond-mat/9708200.  

\item{[8]} Planck, M., 1926 {\it \"Uber die Begrundung des zweiten
Hauptsatzes der Thermodynamik,} Sitzungsber.~Preuss.~Akad. Wiss.,
Phys.~Math.~Kl., 453-463.

\item{[9]} Roberts, F.~S. and Luce, R.~D., 1968, {\it Axiomatic
thermodynamics and extensive measurement}, Synthese {\bf 18}, 311-326.

%%%%%%%%%%%%%

\end